\makeatletter
\@ifundefined{ResumeTagging}{\def\ResumeTagging#1{}}{}
\@ifundefined{SuspendTagging}{\def\SuspendTagging#1{}}{}
\@ifundefined{UseTaggingSocket}{\def\UseTaggingSocket#1{}}{}
\makeatother

\documentclass{SciPost}

\hypersetup{
    colorlinks,
    linkcolor={red!50!black},
    citecolor={blue!50!black},
    urlcolor={blue!80!black}
}

\usepackage[bitstream-charter]{mathdesign}
\DeclareSymbolFont{usualmathcal}{OMS}{cmsy}{m}{n}
\DeclareSymbolFontAlphabet{\mathcal}{usualmathcal}

\fancypagestyle{SPstyle}{
\fancyhf{}
\lhead{\colorbox{scipostblue}{\bf \color{white} ~SciPost Physics Proceedings }}
\rhead{{\bf \color{scipostdeepblue} ~Submission }}

\fancyfoot[C]{\textbf{\thepage}}
}

\newcommand{\rmd}{{\rm d}}
\newcommand{\rmi}{{\rm i}}
\newcommand{\rme}{{\rm e}}
\newcommand{\red}{\color{black}}

\begin{document}

\pagestyle{SPstyle}

\begin{center}{\Large \textbf{\color{scipostdeepblue}{
%%%%%%%%%% TODO: Write your article's title here
Hidden Supersymmetry in Wigner-Yang Quantum Mechanics\footnote{Dedicated to Asim Orhan Barut (1926-1994) on the occasion of his $100^{\rm th}$ birthday.}
%Article Title, as descriptive as possible, ideally fitting in two lines (approximately 150 characters) or less\\
}}}\end{center}

\begin{center}\textbf{
% Use (full) first name (+ middle name initials) + surname format.
% Separate subsequent authors by a comma, omit comma and use "and" for the last author.
% Mark the corresponding author(s) with a superscript symbol in this order
% \star, \dagger, \ddagger, \circ, \S, \P, \parallel, ...
Georg Junker%\textsuperscript{1$\star$},
}\end{center}

\begin{center}
%{\bf 1}
Institut für Theoretische Physik,
Friedrich Alexander Universität Erlangen-Nürnberg,
Staudtstraße 7, 91058 Erlangen, Germany
%%%%%%%%%% END TODO: AFFILIATIONS
% Provide email address of corresponding author(s)
\\[\baselineskip]
%$\star$
\href{mailto:email1}{\small georg.junker@fau.de}
\end{center}

\definecolor{palegray}{gray}{0.95}
\begin{center}
\colorbox{palegray}{
  \begin{tabular}{ll}
 \hspace{-11pt} \begin{minipage}{0.18\textwidth}
    \includegraphics[width=28mm]{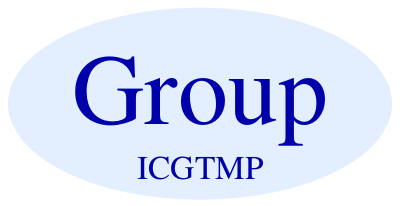}
  \end{minipage}
  &
  \begin{minipage}{0.75\textwidth}
    \begin{center}
    {\it 36th International Colloquium on Group Theoretical Methods in Physics}\\
    {\it Valladolid, 13-17 July 2026} \\
    \doi{10.21468/SciPostPhysProc.?}\\
    \end{center}
  \end{minipage}
\end{tabular}
}
\end{center}

\section*{\color{scipostdeepblue}{Abstract}}
\textbf{\boldmath{%
%%%%%%%%%% TODO: frame
We consider a quantum system on the real line obeying a deformed Heisenberg algebra.
%%-Comment 1
{\red This deformation was originally proposed by E.P.\ Wigner in 1950 when establishing the non-uniqueness of the canonical commutation relations implied by the equations of motion. An explicit coordinate representation was subsequently provided by L.M.\ Yang}
%%+Comment 1
in 1951 and in essence is identical in form with Dunkl's difference-differential operator introduced in 1989 in connection with root systems of reflection groups
%%-Comment 1
{\red when limited to the group $Z_2$ acting on the real line.}
%%+Comment 1
Under certain conditions such quantum systems exhibit a supersymmetric (SUSY) structure where the reflection operator acts as the grading operator. We present a generalisation of Yang's representation by first considering only one of the two equations of motion in phase space. The corresponding non-interacting system is found to represent Witten's model of SUSY quantum mechanics. Imposing also the second equation of motion the original result of Wigner and Yang is reconsidered by extending their discussion to general symmetric potentials on the real line. As explicit example we discuss the harmonic oscillator and an attractive Coulomb-like potential $V(x)=-\gamma/|x|$. We also establish a Hooke-Newton duality between this Coulomb-like system and the original Wigner-Yang harmonic oscillator system.
}}

\vspace{\baselineskip}

%%%%%%%%%% BLOCK: Copyright information
% This block will be filled during the proof stage, and finilized just before publication.
% It exists here only as a placeholder, and should not be modified by authors.
\noindent\textcolor{white!90!black}{%
\fbox{\parbox{0.975\linewidth}{%
\textcolor{white!40!black}{\begin{tabular}{lr}%
  \begin{minipage}{0.6\textwidth}%
    {\small Copyright attribution to authors. \newline
    This work is a submission to SciPost Phys. Proc. \newline
    License information to appear upon publication. \newline
    Publication information to appear upon publication.}
  \end{minipage} & \begin{minipage}{0.4\textwidth}
    {\small Received Date \newline Accepted Date \newline Published Date}%
  \end{minipage}
\end{tabular}}
}}
}
%%%%%%%%%% BLOCK: Copyright information

% For convenience during refereeing we turn on line numbers:
%\linenumbers

\vspace{10pt}
\noindent\rule{\textwidth}{1pt}
\tableofcontents
\noindent\rule{\textwidth}{1pt}
\vspace{10pt}

% Write your article contents here, starting from first \section.
% An example structure is given below.

\section{Introduction}
\label{sec1}
% TODO: write your article here.
In 1950 Eugene Wigner \cite{Wigner1950} asked himself the question: "Do the Equations of Motion Determine the Quantum Mechanical Commutation Relations?". He considered the linear harmonic oscillator Hamiltonian and observed that the canonical commutation relations are not necessarily uniquely determined by the equations of motion. Whereas Wigner only determined an implicit generalisation of the canonical commutation relations, one year later, Yang \cite{Yang1951} obtained an explicit $x$-representation of the momentum operator being of the form
\begin{equation}\label{YangP}
  p = -\rmi \hbar \left(\partial _x - \frac{\nu}{x}\, R \right)
\end{equation}
with $\nu$ being a priori an arbitrary real constant and $R$ stands for the reflection operator on the real line acting on functions $g(x)$ as follows $(Rg)(x)=g(-x)$. The above Yang derivative $\partial _x - \frac{\nu}{x}\, R$ is closely related to the Dunkl operator $\partial _x + \frac{\nu}{x}\,(1-R)$ being a differential-difference operator introduced by Charles Dunkl in 1989 \cite{Dunkl1989}. Actually, the additional term $\frac{\nu}{x}$ present in Dunkl's derivative may always be removed via a kind of gauge transformation as already observed by Yang. Note that Yang's representation results in a deformed quantum mechanics based on the deformed Heisenberg algebra
\begin{equation}\label{YangCom}
  [p,x]= -\rmi\hbar(1+2\nu R)\,,
\end{equation}
with $\nu$ being the deformation parameter. For the harmonic oscillator system being a non-negative Hamiltonian, Yang limited this parameter to values in the range $\nu \ge -\frac{1}{2}$. However, in 1978 Ohnuki and Kamefuchi \cite{OK1978} showed that one may allow for any real $\nu$ when wave functions are considered as so-called hyperfunctions.

Above deformed commutation relation expressed in terms of bosonic or fermionic creation and annihilation operators was then utilised in 1953 by Green \cite{Green1953} in the framework of a generalised field quantisation method. See also
%-Comment 2
{\red the work by Mukunda et al \cite{Mukunda1980} and}
%+Comment 2
the monograph by Ohnuki and Kamefuchi \cite{OK1982}
%-Comment 2
{\red on parastatistics.}
{\red
As a side remark let us briefly mention that parastatistics resolved the quark-parton-paradox and paved the road towards the $SU(3)$ color-degree of freedom of quarks\cite{Greenberg1964}.}
%+Comment 2

Since the 1990s deformed quantum mechanics, where the spatial derivative is replaced by the Yang or Dunkl derivative, has attracted increasing interest in the literature and evolved into a very active area of mathematical physics. One of the pioneers has been Mikhail Plyushchay \cite{Plyushchay1994,Plyushchay1996,Plyushchay1996b,Plyushchay1997,Plyushchay2000}, who studied various aspects of deformed Heisenberg algebra involving reflection operator. Since then Wigner-Yang, respectively, Wigner-Yang-Dunkl quantum mechanics, has attracted much attention. See also the various contributions to this conference.
%-Comment 7
{\red For broader physical relevance of the reflection-deformed algebra beyond one-dimensional quantum mechanics, we refer to the work be Horváthy and Plyushchay \cite{Horvathy2004} and Klishevich et al \cite{Klishevich2001}.
}
%+Comment 7

In this work we will revisit the Wigner-Yang approach, however, for an a priori general but symmetric potential $V(x)=V(-x)$. Implementing the first equation of motion $\dot{p}=-V'(x)$, which in essence reflects Newton's law, we introduce a generalised Wigner-Yang derivative and its associated momentum operator. The corresponding quantum system, called generalised Wigner-Yang quantum mechanics (gWYQM), will be discussed. Considering the reflection operator as a grading operator of the underlying Hilbert space we rediscover Witten's model of supersymmetric quantum mechanics (SUSY QM) as discussed by Plyushchay \cite{Plyushchay1994,Plyushchay1996,Plyushchay1996b} and Beckers et al \cite{Beckers1997}. After the general discussion we implement the second equation of motion $p=\dot{x}$ in Section 3. Here we recall the findings of Wigner and Yang concerning the harmonic oscillator system, nowadays called Wigner quantum oscillator \cite{Palev1994}. We also discuss an attractive Coulomb-like potential exhibiting a SUSY structure. Finally, we establish in Section 4 the Hooke-Newton duality between the harmonic (Hooke) and the Coulomb (Newton) WYQM system. A brief summary with an outlook for further investigations is provided in Section 5.

\section{Generalised Wigner-Yang quantum mechanics}
\label{sec2}
Let us revisit Wigner's question for a generic Hamiltonian on the real line
\begin{equation}\label{Hgen}
  H=\frac{p^2}{2m}+V(x)\,.
\end{equation}
In the above $p$ stands for the linear momentum operator and the potential $V$ is an arbitrary well-defined function of the position operator $x\in\mathbb{R}$. The first equation of motion (EoM) should be identical to its classical counterpart and read
\begin{equation}\label{EoMgen}
  \dot{p}=\frac{\rmi}{\hbar}\, [H,p]=- V'(x)\,.
\end{equation}
Obviously the standard coordinate representation $p=-\rmi\hbar \partial_x$, i.e.\ $[p,x]=-\rmi\hbar$, would reproduce above EoM. However, as pointed out by Wigner and more explicit by Yang, this is not unique. Indeed, for arbitrary but symmetric potentials $V(-x)=V(x)$, Yang's ansatz
\begin{equation}\label{genWYP}
  P:=-\rmi\hbar\Bigl[ \partial_x -f(x) R\Bigr]\,,
\end{equation}
would reproduce above EoM for an arbitrary function $f(x)$ on the real line.
Recall, the reflection operator $R$ acts on arbitrary functions $g$ on the real line as follows, $Rg(x)=g(-x)$.
%-Comment 9
{\red
In particular, $R$ and $\partial_x$ anti-commute with $R$, i.e.\ $\{R,x\}=0$ and $\{R,\partial_x\}= 0$.
}
%+Comment 9
In order to have a self-adjoint momentum operator $P^\dag=P$ we must require that function $f$ is an odd function, i.e.\ $f(-x)=-f(x)$. In other words, for any symmetric potential $V$ and an arbitrary anti-symmetric function $f$, the representation (\ref{genWYP}) is a possible quantisation holding the same EoM as the standard canonical quantisation. We will call quantum dynamics based on the momentum operator represented by (\ref{genWYP}) generalised Wigner-Yang quantum mechanics (gWYQM). The corresponding generalised Wigner-Yang derivative is then given by
\begin{equation}\label{gWYD}
  D_x := \partial_x -f(x)\, R
\end{equation}
and acts on states in Hilbert space ${\cal H}:=L^2(\mathbb{R})$. As a consequence, the commutation relation between $D_x$ and $x$ reads
\begin{equation}\label{gWAlgebra}
  [x,D_x]=1+2xf(x)R\,.
\end{equation}
The standard canonical commutation relation is recovered when setting $f\equiv 0$.
A generic Hamiltonian of gWYQM is given by
\begin{equation}\label{gWYH}
  H:= -\frac{\hbar^2}{2m} D^2_x +V(x)=-\frac{\hbar^2}{2m}\partial_x^2 + \frac{\hbar^2}{2m}f^2(x)+\frac{\hbar^2}{2m}f'(x)R + V(x)\,.
\end{equation}
Note that above Hamiltonian commutes with the parity operator $R$, i.e.\ $[H,R]=0$. Hence, they have a common set of eigenstates and we may consider $R$ as grading operator splitting the Hilbert space into two subsets of even and odd functions, respectively, ${\cal H}={\cal H}^+ \oplus {\cal H}^-$ with
\begin{equation}\label{calH+-}
  {\cal H}^\pm := \left\{ \psi\in{\cal H}|R\psi=\pm\psi\right\}\,.
\end{equation}

At this stage, let us briefly discuss the close relation between the generalised Wigner-Yang derivative and that introduced by Dunkl in 1989, related to the reflection group $Z_2$ on the real line given by
\begin{equation}\label{gWYDD}
  \tilde{D}_x:= \partial_x +g'(x)-f(x)\, R\,.
\end{equation}
Here $g$ is assumed to be an even but otherwise arbitrary function, $g(-x)=g(x)$.
Note that with substitution $\psi=\rme^{-g}\varphi$ one can gauge away this additional term as $\tilde{D}_x\psi =\rme^{-g} D_x\varphi $ and hence, generalised Wigner-Yang-Dunkl quantum mechanics (gWYDQM) with momentum operator $\tilde{p}=-\rmi \hbar \tilde{D}_x$ on $\tilde{\cal H}=L^2(\mathbb{R},\rmd\mu)$ with weighted measure $\rmd \mu(x)= \rme^{2g(x)}\rmd x$ is trivially reduced to gWYQM on ${\cal H}=L^2(\mathbb{R})$. In the original approach Dunkl's choice was $g(x)=\nu\ln(|x|)$ with $\nu>-\frac{1}{2}$ and $f(x)=\frac{\nu}{x}$ as by Yang. In essence, gWYDQM and gWYQM are equivalent and therefore, we will consider only gWYQM in our way forward.

\subsection{The hidden SUSY structure in free gWYQM}

Let us consider gWYQM on Hilbert space ${\cal H}:= L^2(\mathbb{R})$ with an odd but otherwise arbitrary function $f(-x)=-f(x)$ and vanishing potential. The Hamiltonian characterising this free gWYQM is given in eq.\ (\ref{gWYH}) with $V(x)=0$.
\begin{equation}\label{gWYHfree}
  H=\frac{\hbar^2}{2m}\left[-\partial_x^2 + f^2(x)+f'(x)R\right]\,.
\end{equation}
Obviously, above Hamiltonian is identical in form with Witten's model of SUSY QM. See, for example, \cite{Junker2019}. The self-adjoint SUSY charges are given by
\begin{equation}\label{gWYQMQ}
  Q_1:=\frac{\rmi\hbar}{\sqrt{4m}}\,D_x\,,\qquad Q_2:=\rmi R\,Q_1\,,
\end{equation}
and obey the SUSY algebra, ($i,j\in\{1,2\}$)
\begin{equation}\label{SUSYalgebra}
  \left\{Q_i,Q_j\right\}=H\delta_{ij}\,,\qquad Q_i^\dag=Q_i\,,\qquad \left\{R,Q_i\right\}=0\,,\qquad R^2 =1\,,
\end{equation}
with reflection operator $R$ acting as grading or Witten operator. In matrix representation, where $\psi^\pm(-x)=\pm\psi^\pm(x)\in{\cal H}^\pm$, we have
\begin{equation}\label{SUSYmatrixreps}
\begin{array}{c}
  \psi=\left(\begin{array}{c} \psi^+\\ \psi^- \end{array}\right)\,,\qquad
  R=\left(\begin{array}{cc} 1&0\\ 0 &-1\end{array}\right)\,,\qquad
  H=\left(\begin{array}{cc} AA^\dag &0\\ 0 &A^\dag A\end{array}\right)\,, \\[6mm]
  \displaystyle
  Q_1=\frac{1}{\sqrt{2}}\left(\begin{array}{cc}0 &A\\ A^\dag &0\end{array}\right)\,,\qquad
  Q_2=\frac{1}{\sqrt{2}}\left(\begin{array}{cc}0 &\rmi A\\ -\rmi A^\dag &0\end{array}\right)\,.
\end{array}
\end{equation}
In above we introduced operator
\begin{equation}\label{A}
  A:=\frac{\rmi \hbar}{\sqrt{2m}}\Bigl(\partial_x+f(x)\Bigr)
\end{equation}
acting as follows
\begin{equation}\label{Adomain}
  A:{\cal H}^- \to {\cal H}^+\,,\qquad A^\dag : {\cal H}^+ \to {\cal H}^-\,.
\end{equation}
It represents in essence the generalised Wigner-Yang derivative (\ref{gWYD}) restricted to the corresponding subspace.

SUSY is said to be unbroken if the ground state of Hamiltonian $H$ has a vanishing energy eigenvalue $E_0=0$. If $E_0>0$ then SUSY is said to be spontaneously broken. Whether SUSY is broken or not depends on the explicit form of function $f$. Assuming an unbroken SUSY with ground state $\psi_0^-\in{\cal H}^-$ we have
\begin{equation}\label{psi0-}
  A\psi_0^- =0\qquad\Longrightarrow\qquad \psi_0^-(x)=N\exp\left\{-\int^{x}\rmd z\,f(z)\right\}=\psi_0^-(-x)\,.
\end{equation}
As $f$ is an odd function $\psi_0^-$ must be even and, hence, cannot belong to ${\cal H}^-$. On the other hand, assuming  unbroken SUSY with ground state $\psi_0^+\in{\cal H}^+$ we conclude
\begin{equation}\label{psi0+}
  A^\dag\psi_0^+ =0\qquad\Longrightarrow\qquad \psi_0^+(x)=N\exp\left\{\int^{x}\rmd z\,f(z)\right\}=\psi_0^+(-x)\,.
\end{equation}
That is for unbroken SUSY the ground state necessarily is an even function belonging to ${\cal H}^+$ and
$f$ must obey the conditions
\begin{equation}\label{fcond}
  \lim_{x\to-\infty}f(x) > 0 > \lim_{x\to +\infty}f(x)
\end{equation}
for normalisability reasons.\\

\noindent {\it Example for unbroken SUSY:}\\
Let us choose $f(x)=-\frac{m\omega}{\hbar}\,x$ with $\omega>0$ obeying condition \eqref{fcond}. The corresponding WYQM Hamiltonian can be written in terms of the standard harmonic oscillator annihilation and creation operators $a$ and $a^\dag$.
\begin{equation}\label{HOunbroken}
  H=\hbar\omega\left(a^\dag a +\frac{1-R}{2}\right)\,.
\end{equation}
The common eigenstates may be denoted by $|n\rangle$, and obey the relations $a^\dag a|n\rangle=n|n\rangle$ and $R|n\rangle=(-1)^n|n\rangle$, where $n=0,1,2,\ldots$. The corresponding spectrum of $H$ is then given by
\begin{equation}\label{EHO}
  E_n=\hbar\omega \left(n+\frac{1}{2}\Bigl(1-(-1)^n\Bigr)\right).
\end{equation}
SUSY is unbroken as $E_0=0$ and we have the relations $E_{2s}=2\hbar\omega s$ and $E_{2s+1}=2\hbar\omega (s+1)$ indicating the essential isospectrality between the SUSY partner Hamiltonians $H_+=AA^\dag=\hbar\omega a^\dag a$ and $H_-=A^\dag A=\hbar\omega aa^\dag$. Note that $A=-\rmi\sqrt{\hbar\omega}\, a^\dag$ and $A^\dag=\rmi\sqrt{\hbar\omega}\, a$. Hence the SUSY transformations \eqref{Adomain} are trivially found
\begin{equation}\label{SUSYtransHO}
  A|2s-1\rangle =-\rmi\sqrt{E_{2s}}|2s\rangle\,,\qquad  A^\dag|2s\rangle=\rmi\sqrt{E_{2s}}|2s-1\rangle\,,\qquad
  s=1,2,3,\ldots\,.
\end{equation}

\noindent {\it Example for broken SUSY:}\\
Now we chose  $f(x)=\frac{m\omega}{\hbar}\,x$ resulting in a broken SUSY as discussed above. In this case the Hamiltonian reads
\begin{equation}\label{HObroken}
  H=\hbar\omega\left(a^\dag a +\frac{1+R}{2}\right)
\end{equation}
and its eigenvalues are given by
\begin{equation}\label{EHO2}
  E_n=\hbar\omega \left(n+\frac{1}{2}\Bigl(1+(-1)^n\Bigr)\right)\,,
\end{equation}
that is, we have strict isospectrality between even and odd eigenstates, $E_{2s}=\hbar\omega(2s+1)=E_{2s+1}$.

Here we note that both examples were already discussed by Beckers et al \cite{Beckers1997} in a slightly different context. They connected the general case of odd $f$ with its reducibility as representation of $sl(1|1)$ or $sqm(2)$, which fails for the case of an even $f$ requiring them to introduce a third supercharge extending $sqm(2)$ to $sqm(3)$.
%-Comment 4
{\red
Such a third supercharge also occurs, for example, in the fermion-monopole supersymmetry as discussed by Plyushchay in \cite{Plyushchay2000b}.
}
%+Comment 4
%-Comment 3
{\red
For further discussions on the relation between supersymmetric quantum mechanics and its bosonized
realization, being closely related to the present one, let us refer to \cite{Gamboa1999}.
}
%+Comment 3

In concluding this section we can state that the free gWYQM constitutes as substructure of Witten's SUSY QM. As the SUSY potential is an odd function, when SUSY is unbroken, its ground state belongs to ${\cal H}^+$. Here the reflection operator $R$ acts as grading or Witten parity operator. There is, however, an essential difference to the standard Witten model of SUSY QM \cite{Junker2019}. In Witten's model both Hamiltonians act on states in ${\cal H}=L^2(\mathbb{R})$. In the free gWYQM the Hamiltonians $H_\pm$ act on the subspaces ${\cal H}^\pm$ of even and odd functions, respectively. Nevertheless, the operator \eqref{A} generates SUSY transformations according to \eqref{Adomain}.
%-Comment 4
{\red
As a final remark let us mention that the reflection operator is not the only possible grading choice
in systems with additional supercharges  \cite{Correa2008}.
}
%+Comment 4

\section{Wigner-Yang quantum mechanics with interaction}

We will now turn to the standard WYQM as originally proposed by Yang \cite{Yang1951}. In fact, taking now the second EoM $\dot{x}=\frac{\rmi}{\hbar}[H,x]=p$ additionally in consideration, one is forced to particular solution $f(x)=\frac{\nu}{x}$, with an arbitrary real deformation parameter $\nu$. Here, in contrast to Wigner and Yang, we allow for generic non-vanishing symmetric potential $V(-x)=V(x)\neq 0$.
As in the previous section, the Hamiltonian (\ref{gWYH}) commutes with $R$ and we again may split the Hilbert space into subspaces of even and odd functions. The Hamiltonian is then represented by the pair
\begin{equation}\label{HWYpm}
  H^{(\nu)}_\pm := -\frac{\hbar^2}{2m}\partial_x^2 + \frac{\hbar^2\nu(\nu\mp 1)}{2mx^2} + V(x)\qquad\mbox{on}\qquad {\cal H}^\pm= \left\{ \psi\in L^2(\mathbb{R})|R\psi(x)=\pm\psi(x)\right\}\,.
\end{equation}

As the potential is symmetric we may further restrict the eigenvalue problem of $H_\pm$ to the positive half-line $r\geq 0$ with proper boundary conditions at the origin $r=0$. That is, we may consider the restricted pair of Hamiltonians
\begin{equation}\label{radHWYpm}
  \hat{H}^{(\nu)}_\pm = -\frac{\hbar^2}{2m}\partial_r^2 + \frac{\hbar^2\nu(\nu\mp 1)}{2mr^2} + V(r)\qquad\mbox{on}\qquad \hat{\cal H}^\pm\,,
\end{equation}
where the two subspaces are now given by
\begin{equation}\label{hatcalH}
  \hat{\cal H}^+:=\{L^2(\mathbb{R}^+)|\psi' (0)=0 \}\,,\qquad \hat{\cal H}^-:=\{L^2(\mathbb{R}^+)|\psi (0)=0 \}\,.
\end{equation}
That is, we require Neumann and Dirichlet boundary conditions at the origin, respectively. In that way we have formally reduced the eigenvalue problem of $H_\pm$ to that of two standard radial problems (\ref{radHWYpm}) with, in general non-integer, angular momentum $\ell_+:=\nu -1$ and $\ell_-:=\nu$, respectively.  Note however, the different boundary conditions (\ref{hatcalH}) being applied. Here for simplicity we restrict ourselves to Neumann's condition on $\hat{\cal H}^+$. In fact one could also allow for a diverging wave function near $r=0$, which may still be square integrable.
%-Comment 8
{\red
Self-adjointness of Hamiltonians and related supercharges clearly require a careful discussion of their respective domains. However, an extensive functional-analytic discussion is beyond the scope of this work. In the current context we will limit ourselves to above boundary conditions and further restrictions as presented below.
}
%% +Comment 8

Let us look into the eigenvalue problem of the radial Hamiltonians, assuming a purely discrete spectrum
\begin{equation}\label{HpmEV}
  \hat{H}^{(\nu)}_\pm \, \psi_n^\pm = E_{n}^\pm\psi_n^\pm\,,\qquad n=0,1,2,3,\ldots\,.
\end{equation}
We further restrict the potential $V(r)$ to behave near the origin $r\to 0$ like $V(r)\sim ar^\gamma$ with exponent  $\gamma > -2$. Under that assumption eq.\ (\ref{HpmEV}) is dominated by the "centrifugal" term and we conclude the eigenfunctions near the origin must behave like
\begin{equation}\label{psipm}
  \psi_n^+ \sim r^\nu\,,\qquad \psi_n^- \sim r^{\nu-1}\,.
\end{equation}
In both cases the respective boundary conditions require us to the restriction $\nu > 1$. In other words, for well-behaved potentials as assumed here and with $\nu > 1$ both, Neumann and Dirichlet boundary conditions, are automatically obeyed in both subspaces. In essence we may set $\hat{\cal H}^+ = \hat{\cal H}^- = L^2(\mathbb{R}^+)$ and consider the eigenvalue problem (\ref{HpmEV}) on $L^2(\mathbb{R}^+)$. That way, the WYQM eigenvalue problem (\ref{HWYpm}) on the full line $x\in\mathbb{R}$ may be reduced to a standard radial eigenvalue problem \eqref{radHWYpm} on the positive half-line $r\in\mathbb{R}^+$. The eigenfunctions of the latter are then analytically continued to the negative half-line via $\psi^\pm(-r)=\pm\psi^\pm(r)$.
%-Comment 8
{\red
A posteriori, one may then relax above conditions on the deformation parameter $\nu$ as appropriate.
}
%% +Comment 8

By looking at  eq.\ \eqref{HpmEV} we observe that $\hat{H}^{(\nu)}_+=\hat{H}^{(\nu-1)}_-$ and conclude  that $E_n^+(\nu)= E_{n}^{-}(\nu-1)$. This kind of degeneracy between different WYQM systems with angular momentum parameter $\ell_-:=\nu$ and $\ell_+:=\nu-1$ can be a signature of a hidden supersymmetry. Note the close similarity between the WYQM Hamiltonian (\ref{radHWYpm}) and the one for the Pauli Hamiltonian interacting with a spherically symmetric scalar potential exhibiting SUSY \cite{Junker2025}. There, when restricting the Pauli problem to a subspace with fixed total angular momentum, the two SUSY partner Hamiltonian also contain a centrifugal potential with orbital angular momentum $\ell -1$ and $\ell$, respectively.

\subsection{Example 1: The harmonic oscillator (Hooke's law)}
As a first example let us consider the harmonic interaction
\begin{equation}\label{HO}
  V(x)=\frac{m}{2}\omega^2 x^2\,,\qquad\omega >0\,,
\end{equation}
which basically reduces (\ref{HpmEV}) to the eigenvalue problem of the radial oscillator, a standard textbook problem in quantum mechanics,%-Comment 5
\footnote{\red
Such Hamiltonians are closely related to the conformal mechanics of de Alfaro, Fubini and Furlan \cite{Alfaro1976}.
}
%+Comment 5
with fixed angular momentum $\ell_+=\nu -1$ and $\ell_-=\nu$.
\begin{equation}\label{HHO}
  H_H^{(\nu)}:=  -\frac{\hbar^2}{2m}\partial_x^2 + \frac{\hbar^2\nu(\nu -R)}{2mx^2} + \frac{m}{2}\omega^2 x^2\,.
\end{equation}
The well-known results are the eigenvalues
\begin{equation}\label{HOEV}
\textstyle
  E_{n}^+=\hbar\omega(2n +\nu +\frac{1}{2})\,,\qquad E_{n}^-=\hbar\omega(2n +\nu +\frac{3}{2})\,.
\end{equation}
with associated eigenfunctions $(\alpha^2:=m\omega/\hbar)$
\begin{equation}\label{HOEF}
\begin{array}{rcl}
  \psi^+_{n}(x) & = & \displaystyle\sqrt{\frac{\alpha\,\Gamma(n+1)}{\Gamma(n+\nu+\frac{1}{2})}}\,(\alpha^2 x^2)^{\nu/2}\rme^{-\alpha^2 x^2/2} L^{\nu-\frac{1}{2}}_{n}(\alpha^2 x^2)\,,\\[6mm]
   \psi^-_{n}(x) & = & \displaystyle\sqrt{\frac{\alpha\,\Gamma(n +1)}{\Gamma(n+\nu+\frac{3}{2})}}\,{\rm sgn\,}(x)(\alpha^2 x^2)^{(\nu+1)/2}\rme^{-\alpha^2 x^2/2} L^{\nu+\frac{1}{2}}_{n}(\alpha^2 x^2)\,.
\end{array}
\end{equation}
Note that above eigenfunctions are normalised on ${\cal H}^\pm$. Obviously, the spectrum of the Wigner-Yang harmonic oscillator $E_{n}=\hbar\omega(n+\nu+\frac{1}{2})$ is recovered with $E_{2n}=E_{n}^+$ and $E_{2n+1}=E_{n}^-$ and similarly for the eigenfunctions. Clearly the set of eigenvalues for the even and odd solutions are disjoint and cannot result in a SUSY structure. It is the non-local harmonic potential $\tilde{V}(x)=\frac{m}{2}\omega^2 x^2+\frac{\hbar\omega}{2}\,R$, which allows for a SUSY structure. Naturally we ask if there exists a local potential exhibiting such a SUSY structure. This is the objective of the next example.
\subsection{Example 2: The Coulomb-like potential (Newton's law)}
As second example we consider the Coulomb-like attractive potential
\begin{equation}\label{Cou}
  V(x)=-\frac{\gamma}{|x|}\,,\qquad\gamma >0\,,
\end{equation}
resulting in the pair of radial Hamiltonians
\begin{equation}\label{HradCou}
  \hat{H}^{(\nu)}_\pm = -\frac{\hbar^2}{2m}\partial_r^2 + \frac{\hbar^2\nu(\nu\mp 1)}{2mr^2} -\frac{\gamma}{r}\,.
\end{equation}
Again the corresponding eigenvalue problem is that of the radial part of the hydrogen atom with fixed angular momentum. This is another well-studied textbook problem resulting in eigenvalues and eigenfunctions as follows.
\begin{equation}\label{CouEV}
  E_{n}^+=-\frac{m\gamma^2}{2\hbar^2(n + \nu)^2}=-\frac{\hbar^2\kappa^2_+}{2m}\,,\qquad E_{n}^-=-\frac{m\gamma^2}{2\hbar^2(n + \nu + 1)^2}=-\frac{\hbar^2\kappa^2_-}{2m}\,.
\end{equation}
\begin{equation}\label{CouEF}
\begin{array}{rcl}
  \psi^+_{n}(x) & = & \displaystyle
  \sqrt{\frac{\kappa_+\,\Gamma(n +1)}{\Gamma(n+2\nu)(2n+2\nu)}}\,(2\kappa_+ |x|)^{\nu}\rme^{-\kappa_+ |x|} L^{2\nu-1}_{n}(2\kappa_+ |x|)\,,\\[6mm]
  \psi^-_{n}(x) & = & \displaystyle
  \sqrt{\frac{\kappa_-\,\Gamma(n +1)}{\Gamma(n +2\nu+2)(2n+2\nu+2)}}\,{\rm sgn\,}(x)(2\kappa_- |x|)^{\nu+1}\rme^{-\kappa_- |x|} L^{2\nu+1}_{n}(2\kappa_- |x|)\,.
\end{array}
\end{equation}
In the above we have set $\kappa_+:=\frac{m\gamma}{\hbar^2(n+\nu)}$ and  $\kappa_-:=\frac{m\gamma}{\hbar^2(n+\nu+1)}$.
Here we note the degeneracy
\begin{equation}\label{CouESUSY}
E_{n +1}^+=E_{n}^-\,,
\end{equation}
which allows for an unbroken SUSY structure of the Coulomb like problem in WYQM. This was already observed by Beckers and Debergh \cite{Beckers1993} who required $\nu>0$ assuring the spectrum being bounded from below. However, in \cite{Beckers1993} the discussion was exclusively limited to the positive half-line omitting the peculiarities when extending the discussion to the full line. We will analyse this SUSY structure in more detail in the next subsection.
\subsection{Hidden SUSY in WYQM with Coulomb-like interaction}
Let us start by introducing ladder operators
\begin{equation}\label{ladderopsrad}
  a_\nu :=\frac{\hbar}{\sqrt{2m}}\left( -\partial_r -\frac{\nu}{r}+\frac{m\gamma}{\hbar^2\nu}\right)\,,\qquad
  a^\dag_\nu =\frac{\hbar}{\sqrt{2m}}\left( \partial_r -\frac{\nu}{r}+\frac{m\gamma}{\hbar^2\nu}\right)\,,
\end{equation}
leading us to the well-known factorisation of the radial Hamiltonians \eqref{HradCou} of the form \cite{Junker2019}
\begin{equation}\label{HradCoupm}
  \hat{H}^{(\nu)}_+ = a_\nu a^\dag_\nu - \frac{m\gamma^2}{2\hbar^2\nu^2}\,,\qquad \hat{H}^{(\nu)}_- = a^\dag_\nu a_\nu - \frac{m\gamma^2}{2\hbar^2\nu^2}\,,
\end{equation}
exhibiting supersymmetry. The question arises, can this be extended to the full line on which the Coulomb-like WYQM system is defined. For this we generalise above ladder operators by setting
\begin{equation}\label{ladderops}
  A_\nu :=\frac{\hbar}{\sqrt{2m}}\left( -\partial_x -\frac{\nu}{x}+\frac{m\gamma}{\hbar^2\nu}\,{\rm sgn\,}x\right)\,,\qquad
  A^\dag_\nu =\frac{\hbar}{\sqrt{2m}}\left( \partial_x -\frac{\nu}{x}+\frac{m\gamma}{\hbar^2\nu} \,{\rm sgn\,}x\right)\,,
\end{equation}
where ${\rm sgn\,}x=\frac{x}{|x|}$. We observe that
\begin{equation}\label{Afactor}
  A^\dag_\nu A_\nu = H^{(\nu)}_- + \frac{m\gamma^2}{2\hbar^2\nu^2} + \frac{\gamma}{2\nu}\,\delta(x) \,,\qquad
  A_\nu A^\dag_\nu = H^{(\nu)}_+ + \frac{m\gamma^2}{2\hbar^2\nu^2} - \frac{\gamma}{2\nu}\,\delta(x) \,.
\end{equation}
The SUSY structure is then explicated by introducing
\begin{equation}\label{HSUSY}
  H_{SUSY}:=\left(\begin{array}{cc} A_\nu A^\dag_\nu & 0 \\ 0 & A^\dag_\nu A_\nu \end{array}\right)=
  %\left(\begin{array}{cc} H^{(\nu)}_+ & 0 \\ 0 &H^{(\nu)}_- \end{array}\right)
  H_N^{(\nu)} +  \frac{m\gamma^2}{2\hbar^2\nu^2}
  -\frac{\gamma}{2\nu}\,\delta(x) R\,.
\end{equation}
Note that in the current context
\begin{equation}\label{Hcoul}
 H_N^{(\nu)} := -\frac{\hbar^2}{2m}\partial_x^2 + \frac{\hbar^2\nu(\nu -R)}{2mx^2} - \frac{\gamma}{|x|}
\end{equation}
is the Coulomb-like WDQM Hamiltonian.  In the SUSY Hamiltonian \eqref{HSUSY} appears besides the trivial constant $\frac{m\gamma^2}{2\hbar^2\nu^2}$ an additional non-local interaction term of the form $\frac{\gamma}{\nu}\,\delta(x) R$. This latter term, however, may be ignored in the eigenvalue problem as according to \eqref{psipm} all eigenfunctions vanish at the origin and hence $\delta(x) R \psi^\pm(x)\equiv 0$.
%-Comment 6
{\red
However, see for example ref.\ \cite{Junker2019,Correa2007}, where such delta-like point interactions are discussed.
}
%+Comment 6

Here it would be sufficient to limit the deformation parameter to $\nu>0$. Hence, from now of we set $H_{SUSY}= H_N^{(\nu)} +  \frac{m\gamma^2}{2\hbar^2\nu^2}$ without loss of generality and establish the SUSY algebra
\begin{equation}\label{SUSYalgebraCoulomb}
   H_{SUSY} =\{Q,Q^\dag\}\,,\qquad Q^2 = 0 = {Q^\dag}^2\,,\qquad [H_{SUSY},R]=0=\{Q,R\}\,,\qquad R^2 =1\,,
\end{equation}
where supercharge and reflection operator are represented by the matrices
\begin{equation}\label{Q}
  Q:= \left(\begin{array}{cc} 0 & A_\nu \\ 0 & 0 \end{array}\right)\qquad \mbox{and}  \qquad
  R:= \left(\begin{array}{cc} 1 & 0 \\ 0 & -1 \end{array}\right)\,,
\end{equation}
respectively. As expected $R$ takes the role of the grading or Witten operator. Here SUSY is unbroken. The zero-energy ground state of $H_{SUSY}$ belongs to ${\cal H}^+$ and is obtained via $A^\dag_\nu \psi^+_{0}=0$ resulting in
\begin{equation}\label{psi0}
  \psi^+_{0}(x)=\left(\frac{m\gamma}{\hbar^2\nu}\right)^{\nu+\frac{1}{2}}  \frac{(2|x|)^\nu}{\sqrt{\Gamma(2\nu+1)}} \exp\left\{-\frac{m\gamma}{\hbar^2\nu}|x|\right\}\,,
\end{equation}
which is indeed consistent with eq.\ \eqref{CouEF}. We may also establish the SUSY transformations utilising the well-known transformation of the standard radial Coulomb problem. First we note that $A^\dag_\nu$ maps an even function into an odd function and vice versa  $A_\nu$ maps an odd function into an even function. More precisely, we can reduce the action of these ladder operators on the full line $x\in\mathbb{R}$ (excluding the origin) into the action of the "radial" ladder operators \eqref{ladderops} on the positive half-line $r=|x|$ by noting that
\begin{equation}\label{Atoa}
  A_\nu (x) = {\rm sgn\,}(x)\, a_\nu(r)\,,\qquad A^\dag_\nu (x) = {\rm sgn\,}(x)\, a^\dag_\nu(r)
\end{equation}
the SUSY transformations read as follows
\begin{equation}\label{SUSYTrans}
\textstyle
  A_\nu \psi^-_n = \sqrt{E_n^- +\frac{m\gamma^2}{2\nu^2}}\,\psi^-_{n+1}\,,\qquad
  A_\nu^\dag \psi^+_n = \sqrt{E_n^+ +\frac{m\gamma^2}{2\nu^2}}\,\psi^+_{n-1}\,,\qquad
  A_\nu^\dag \psi^+_0 = 0\,.
\end{equation}
\section{The Hooke-Newton duality in WYQM}
According to Chandrasekhar's reading of Newton's Principia \cite{Chandr1995}, since the times of Hooke and Newton in the late $17^{\rm th}$ century it is known that the orbits following Hooke's linear force law of elastic springs \cite{Hooke1678} can be mapped into Kepler orbits following Newton's law of gravitation \cite{Newton1687}. This is the so-called Hooke-Newton duality and has been generalised to a broader power-law duality by several authors. For a recent review see \cite{IJ2021}. This duality has also been found helpful in the corresponding quantum problems. For example, the quantum harmonic oscillator eigenvalue problem can be mapped into the eigenvalue problem for the Coulomb interaction, both being the quantum versions of Hooke's and Newton's law, respectively. Having discussed in the previous sections both of these problems within the WYQM framework, it is naturally to investigate this Hooke-Newton duality within the WYQM  formalism. That is, we will try to map the Coulomb-like eigenvalue problem in WYQM
\begin{equation}\label{HCouE}
 \left( H^{(\nu)}_N -E_\nu \right) \psi_N(x)=0
\end{equation}
into the corresponding problem for the harmonic oscillator. In doing so, we consider a change of variables $x\mapsto z$ mapping $\mathbb{R}\to\mathbb{R}$, which we will call modified Hooke-Newton map, as follows
\begin{equation}\label{trans}
  x={\rm sgn\,}(z)\,z^2 = z|z|
\end{equation}
and observe $\partial^2_x  =  \frac{1}{4z^2}\left( \partial_z^2-\frac{1}{z}\,\partial_z \right) $
resulting in
\begin{equation}\label{Delta2}
  D_x^2=\frac{1}{4z^2} \left[\partial_z^2 -\frac{1}{z}\,\partial_z - \frac{4\nu(\nu-R)}{z^2}\right]\,,
\end{equation}
which is not in the form of a squared Wigner-Yang derivative.  Hence we accompany this point transformation with a change of the wave function as follows
\begin{equation}\label{psi}
  \psi_N(x)=: |z|^\alpha \psi_H(z)
\end{equation}
with $\alpha\in\mathbb{R}$ chosen such that we obtain the desired result. Noting that
\begin{equation}\label{D2psi}
  \partial^2_z\psi_N(x)=|z|^\alpha \left[\partial_z^2 +\frac{2\alpha}{z}\partial_z +\frac{\alpha(\alpha-1)}{z^2} \right]\psi_H(z)
\end{equation}
the obvious choice $\alpha=\frac{1}{2}$ removes the term being linear in  the derivative and we finally obtain
\begin{equation}\label{D2phi}
  D_x^2\psi_N(x)= \frac{\sqrt{|z|}}{4z^2}\left[\partial_z^2- \frac{4\nu(\nu-R)+\frac{3}{4}}{z^2}\right]\psi_H(z)\,.
\end{equation}
Here we note that the reflection operator acts as follows
\begin{equation}\label{R}
  R\psi_N(x)=\psi_N(-x)=|z|^\alpha \psi_H(-z)= |z|^\alpha R \psi_H(z)\,.
\end{equation}
As a result we have the relation
\begin{equation}\label{D2phi2}
  D_x^2\psi_N(x)= \frac{\sqrt{|z|}}{4z^2}\left[\partial_z^2- \frac{\mu(\mu-R)}{z^2}\right]\psi_H(z)
\end{equation}
where we formally set
\begin{equation}\label{mu}
  \mu:=2\nu-\frac{1}{2}\,R
\end{equation}
which obviously obeys the relation $\mu(\mu-R)=4\nu(\nu-R)+3/4$. Here we must note that the relation between the deformation parameters $\nu$ and $\mu$ is not an algebraic one but contains the reflection operator.
However, if we restrict ourselves to the subspace with even and odd functions, where $R$ takes the value $s=+1$ and $s=-1$,  respectively, we arrive at the relation
\begin{equation}\label{mus}
  \mu_s = 2\nu - \frac{s}{2}.
\end{equation}
That is, the subspaces ${\cal H}^\pm $ are mapped into each other with different deformation parameters $\mu_+=2\nu-\frac{1}{2}$ and $\mu_-=2\nu+\frac{1}{2}$.

Let us now consider subspaces with fixed parity $s$. Then the above eigenvalue problem \eqref{HCouE} can be put into the form
\begin{equation}\label{map}
 0= \left( H^{(\nu)}_N -E_\nu \right) \psi_N(x)= \frac{\sqrt{|z|}}{4z^2}\left(H^{(\mu_s)}_H-E_{\mu_s} \right)\psi_H(z)\,,
\end{equation}
where we may identify
\begin{equation}\label{Ecoupl}
  E_{\mu_s}=4\gamma\qquad\mbox{and}\qquad\omega^2 = -8E_\nu/m\,.
\end{equation}
The parameter $\mu_s$ is explicitly given by relation \eqref{mus}. Hence the eigenvalue problem for the Coulomb-like potential in WYQM with deformation parameter $\nu$ is reduced to the harmonic oscillator problem in WYQM with deformation parameter $\mu_\pm=2\nu\mp\frac{1}{2}$ restricted to the subspace of even and odd eigenfunction, respectively.

From the harmonic oscillator eigenvalues and eigenfunctions, as given in Section 3.1 above, directly follow those for the Coulomb-like problem and vice versa. Here we only consider the eigenvalues which read
\begin{equation}\label{HOeigen}
\begin{array}{l}
 E_{\mu_+}= \hbar\omega (2n + \mu_+ + \frac{1}{2})= 2\hbar\omega_+ (n+\nu)\,, \\
 E_{\mu_-}= \hbar\omega (2n + \mu_- + \frac{3}{2})= 2\hbar\omega_-(n+\nu +1)\,.
\end{array}
\end{equation}
Utilizing the relations \eqref{Ecoupl}, note that we replaced $\omega$ by $\omega_\pm$ to separate the two subspaces with different index $\mu_\pm$, we directly obtain the Coulomb eigenvalues
\begin{equation}\label{Couleigen}
 E_\nu^+= -\frac{m\omega_+^2}{8}=-\frac{m\gamma^2}{2\hbar^2(n + \nu)^2} \,,\qquad
 E_\nu^-= -\frac{m\omega_-^2}{8}=-\frac{m\gamma^2}{2\hbar^2(n + \nu +1)^2} \,,
\end{equation}
which is indeed the result obtained before. Using the relation $\psi^\pm_N(x)=\sqrt{|z|}\psi^\pm_H(z)$ in combination with $m\omega_\pm=2\hbar\kappa_\pm$ we may verify that the harmonic oscillator eigenfunctions $\psi^\pm_H$ as given in  \eqref{HOEF} transform into the Coulomb eigenfunctions $\psi^\pm_N$ of eq.\ \eqref{CouEF}, except for the normalization factor.

\section{Summary and outlook}
In this presentation we revisited Wigner's question on the uniqueness of the canonical commutation relations. We first considered only one EoM resulting in a generalised Wigner-Yang derivative \eqref{gWYD} leading us to gWYQM. Here we note that for the special case where $f(x)=$ $(2\alpha+1)\coth x+(2\beta+1)\tanh x$ this derivative is called Jacobi-Dunkl operator introduced by Chouchane et al \cite{Chou2003} as a hyperbolic analogue of Dunkl's operator with various applications. We considered  free gWYQM and highlighted its close relation to Witten's model. It was shown how the asymptotic behaviour of the deformation function $f(x)$ rules the SUSY breaking extending previous studies \cite{Plyushchay1996,Plyushchay1997,Beckers1997,Post2011}. We then looked at the standard WYQM and investigated possible  SUSY structures. The general discussion in Section 3 indicates that whenever the effective radial problem results in eigenvalues depending only on the sum $n+\nu$ of radial quantum number and deformation parameter, there is a hidden SUSY. The well-known example of a Coulomb-like potential was already observed before \cite{Beckers1993}. Here we extent the discussion by considering the problem on the full real line and provided explicit expression of the associated SUSY transformation. This may be extended to all conditionally exactly solvable partners of the hydrogen atom problem \cite{JR1998}. We were also able to establish a modified version of the Hooke-Newton duality mapping the Coulomb-like problem in WYQM onto the harmonic oscillator of WYQM. Here we note that relation \eqref{Ecoupl} is identical in form with the usual mapping between energies and coupling constants \cite{IJ2021}. However, the scaling relation of the Langer-modified angular momentum quantum numbers, which reads $\ell_H+\frac{1}{2}= 2(\ell_N+\frac{1}{2})$,  has to be replaced by \eqref{mu}. That scaling relation may be written in the form
\begin{equation}\label{Langer}
  \mu-\frac{R}{2}=2\left(\nu-\frac{R}{2}\right)
\end{equation}
generalising the previous findings \cite{IJ2021}. This Langer-type modification may also be applied to WKB approximations in WYQM. This, as well as, the general power-law duality in WYQM system will be the subject of future investigations.

%\section*{Acknowledgements}
%Acknowledgements should follow immediately after the conclusion.

% TODO: include funding information
\paragraph{Funding information}
The author received no financial support for the research, authorship, and/or publication of this article.


% Provide your bibliography here. You have two options:

%%% FIRST OPTION
% Write your entries here directly, following the example below, including:
% Author(s), Title, Journal Ref. with year in parentheses at the end, followed by the DOI number.

% \begin{thebibliography}{99}
% \bibitem{1931_Bethe_ZP_71}
% H. A. Bethe, \textit{Zur Theorie der Metalle. i. Eigenwerte und Eigenfunktionen der linearen Atomkette}, Zeit. f{\"u}r Phys. \textbf{71}, 205 (1931), \doi{10.1007\%2FBF01341708}.

% \bibitem{arXiv:1108.2700}
% P. Ginsparg, \textit{It was twenty years ago today...}, (arXiv preprint) \doi{10.48550/arXiv.1108.2700}.

% \end{thebibliography}

% Use your bibtex library, formatted by the SciPost style file.
%\bibliography{SciPost_Example_BiBTeX_File.bib}

%%%%%%%%%% END TODO: BIBLIOGRAPHY

\end{document}